\documentclass[prb,twocolumn,showpacs,citeautoscript]{revtex4}
\newcommand{\tablesmallfont}{}

\usepackage{graphicx}
\usepackage{dcolumn}

\newcommand{\WN}{\mbox{cm$^{-1}$}}

\newcommand{\MB}{MgB$_{2}$}
\newcommand{\AB}{AlB$_{2}$}
\newcommand{\etal}{\textit{et al.}}
\newcommand{\Tc}{\ensuremath{T_c}}
\renewcommand{\d}{\mathrm{d}}

\begin{document}

\preprint{\today}

\title{Crystal structure and lattice dynamics of AlB$_2$ under pressure\\ and implications for \MB}
   \author{I. Loa}\email{I.Loa@fkf.mpg.de}
   \author{K. Kunc}
      \altaffiliation[Permanent address: ]{CNRS and Universit\'e P. et M. Curie,
      Laboratoire d'Optique des Solides UMR7601, T13 - C80, 4 pl. Jussieu,
      75252 Paris - C\'edex 05, France.}
   \author{K. Syassen}
   \affiliation{Max-Planck-Institut f\"ur Festk{\"o}rperforschung,
               Heisenbergstrasse 1, D-70569 Stuttgart, Germany}
   \author{P. Bouvier}
   \affiliation{European Synchrotron Radiation Facility,
        BP 220, F-38043 Grenoble, France}

\date{\today}

\begin{abstract}
   The effect of high pressures to 40~GPa on the crystal structure and
   lattice dynamics of \AB\ was studied by synchrotron x-ray powder diffraction,
   Raman spectroscopy, and first-principles calculations. There are no
   indications for a pressure-induced structural phase transition. The
   Raman spectra of the metallic sample exhibit a well-defined peak near
   980~\WN\ at 0~GPa which can be attributed to the Raman-active $E_{2g}$
   zone-center phonon.  Al deficiency of $\sim$11\% in \AB,
   as indicated by the x-ray data, changes qualitatively the electronic
   structure, and there are indications that it may have a sizable effect on
   the pressure dependence of the $E_{2g}$ phonon frequency. Similar
   changes of the pressure dependence of phonon frequencies, caused by
   non-stoichiometry, are proposed as
   an explanation for the unusually large variation of the
   pressure dependence of \Tc\ for different samples of \MB.
\end{abstract}


\maketitle

\section{Introduction}

\begin{figure}[t]
     \centering
     \includegraphics[width=0.8\hsize,clip]{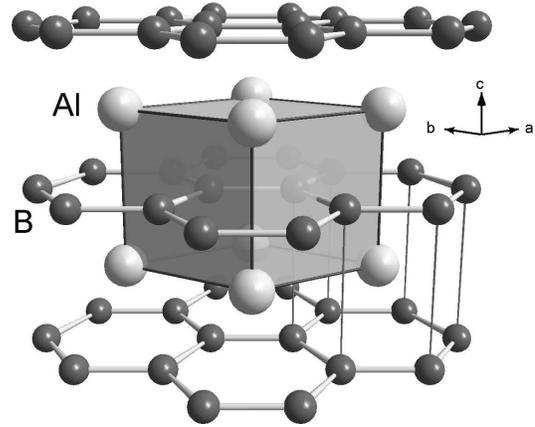}
     \caption{Crystal structure of \AB. The boron atoms form honeycomb
     layers. Al atoms are located at the centers of hexagonal prisms that
     are formed by the B sheets.}
     \label{fig:xtal}
\end{figure}

The \AB\ structure type and derivatives thereof are among the most
frequently occurring ones for intermetallic binary and ternary compounds
\cite{VC97,HP01}.  Transition metal diborides, belonging to this family,
have been studied in some detail because of their potential application in
electronic devices \cite{Will97} to overcome current problems of
electromigration, corrosion, and diffusion into the semiconductor
substrate. The largest interest, however, has undoubtedly received the
recently discovered superconductor \MB\ which also crystallizes in the
simple \AB\ structure depicted in Fig.~\ref{fig:xtal}.

\AB-type compounds have not been studied systematically at high pressures.
There seem to exist no confirmed reports on pressure-induced structural
phase transitions in metal diborides. There are, however, structural
studies of the rare-earth metal digallides GdGa$_{2}$, HoGa$_{2}$,
ErGa$_{2}$, and TmGa$_{2}$ at high pressures
\cite{SBBS01,SBGS98,BSS02p,SBGS96}. With the lighter rare earth elements
La---Er the digallides crystallize in the \AB\ structure at ambient
pressure. In essence, all of the rare-earth metal digallides studied so
far show a transition to the UHg$_{2}$ structure which is isotypic to \AB\
but with a lower $c/a$ ratio.

The superconductor \MB\ was studied at high pressures with regard to its
superconducting transition temperature
\cite{STII01,MNRR01,LMC01,THSH01,TNKK01,LMC01ap}, crystal structure
\cite{VSHY01,PIIC01,JHS01,GSGH01,BMNM01}, and lattice dynamics
\cite{GSGH01,KLSK01}. The pressure dependence of \Tc\ could well be
explained in the framework of phonon-mediated, i.e.\ BCS,
superconductivity \cite{LS01,VSHY01,CZH02}. An isostructural transition
near 30~GPa was reported \cite{LKQZ01} but could not be reproduced in
another study up to 40~GPa \cite{BMNM01}. Much higher pressures may be
necessary to induce transitions possibly towards the UHg$_{2}$ structure.

We study here the effect of hydrostatic pressure on the crystal structure
and lattice dynamics of \AB. Synchrotron x-ray powder diffraction and
Raman scattering experiments are complemented by first-principles
calculations. The present high-pressure study was in part motivated by the
question whether some of the unusual physical properties of \MB\  -- such
as the sizable anharmonicity \cite{YGLB01,KLSK01,LMK01} or the remarkably
large calculated pressure dependence \cite{KLSK01} of the $E_{2g}$ phonon
-- are specific to \MB\ or whether they are characteristic of  other
\AB-type metal diborides as well. It aims at providing high-pressure
structural and lattice dynamical information for comparison with
corresponding data on \MB. Finally, Al deficiency appears to be hardly
avoidable in the growth of \AB\ \cite{MEG64,BGSB00}. We show that it has
significant effect on the electronic structure of \AB\ and may also
influence its lattice dynamics. We will discuss possible consequences of
metal deficiency for the superconductor \MB\ where this issue is also of
relevance.

\section{Experiments}
\label{sec:exp}

\subsection{Experimental Details}

The structural properties of \AB\  under pressure were studied up to 40~GPa
by monochromatic ($\lambda = 0.3738$~{\AA}) x-ray powder diffraction at the
European Synchrotron Radiation Facility (ESRF Grenoble, beamline ID30).
Commercially available \AB\ powder (Alfa Aesar, 99\%) was placed in a
diamond anvil cell (DAC) for pressure generation. Nitrogen was employed as a
pressure medium to provide nearly hydrostatic conditions. Diffraction
patterns were recorded on image plates and then integrated \cite{soft:fit2d}
to yield intensity vs.\ $2\theta$\/ diagrams.

Raman spectra of \AB\ up to 25~GPa (DAC, 4:1~methanol/ethanol mixture as a
pressure medium) were excited at 633~nm utilizing a long-distance microscope
objective. They were recorded in back-scattering geometry using a
single-grating spectrometer with a multi-channel CCD detector and a
holographic notch filter for suppression of the laser line (Dilor Labram).
For the Raman experiments the DAC was equipped with synthetic diamonds
(Sumitomo type IIa) which emit only minimal luminescence. In all experiments
pressures were measured with the ruby luminescence method \cite{MXB86}.

\subsection{X-ray diffraction under pressure}

\begin{figure}
     \centering
     \includegraphics[width=0.7\hsize,clip]{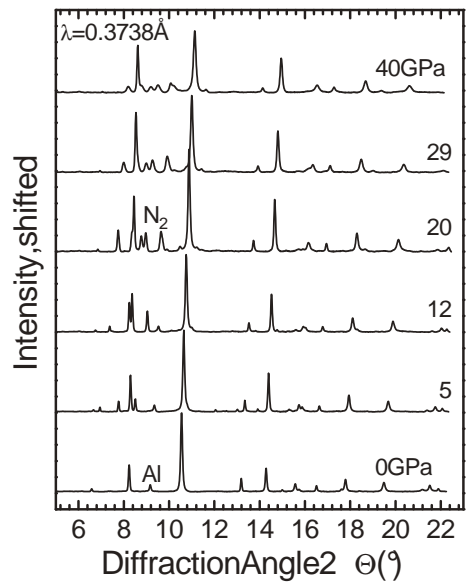}
     \caption{X-ray powder diffraction diagrams of \AB\ at various
     pressures ($T=298$~K). `Al' marks a peak due to fcc-Al.
     Additional peaks appear at pressures above 2~GPa due to various phases of solid
     nitrogen.}
     \label{fig:diffpat}
\end{figure}

Figure~\ref{fig:diffpat} shows x-ray diffraction patterns of \AB\ for
increasing pressures up to 40~GPa. The diagrams evidence small amounts of Al
metal as a secondary phase. At pressures above 2~GPa additional reflections
are observed due to various phases of solid nitrogen. There are no
indications for a pressure-induced structural phase transition in \AB\ up to
40~GPa.

\begin{figure}
     \centering
     \includegraphics[width=\hsize,clip]{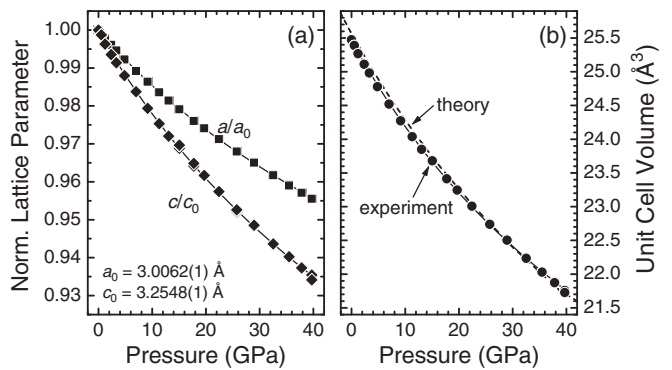}
     \caption{(a) Experimental lattice parameters of \AB\ as a function of
     pressure, normalized to their respective zero-pressure
     values. (b) Experimental (symbols and solid line) and calculated
     (dashed line) pressure--volume relations. The lines are given by the
     Murnaghan equation of state with the parameters listed in
     Table~\ref{tab:structure}.}
     \label{fig:LP-Vol}
\end{figure}

Lattice parameters as a function of pressure were determined from
Rietveld-type fits of the diffraction diagrams. The compressibility of
\AB\ is moderately anisotropic as illustrated in Fig.~\ref{fig:LP-Vol}(a)
with the softer direction being parallel to the $c$ axis. Up to 40~GPa,
compression along $c$ is 47\% larger than along $a$. The $c/a$ ratio
decreases from 1.083 (0~GPa) to 1.060 at 40~GPa. From the lattice
parameters we determine the unit cell volume as a function of pressure as
shown in Fig.~\ref{fig:LP-Vol}(b). The data are well represented by the
Murnaghan relation \cite{Mur44} $V(P) = V_0 [(B'/B_0) P+1]^{-1/B'}$. With
$V_0 = 25.4734(5)$~{\AA}$^3$ fixed at the value determined from the
zero-pressure data we obtain by least-squares fitting the bulk modulus
$B_0$ and its pressure derivative $B'$ at zero pressure as listed in
Table~\ref{tab:structure}.

\begin{table*}
\caption{Structural parameters of \AB\ and \MB: zero-pressure volume $V_0$
and lattice constants $a_0, c_0$; bulk modulus $B_0$, and its pressure
derivative $B'$ at zero pressure. Variation of the $c/a$ ratio with
pressure is described by the quadratic polynomial $c/a = c_0/a_0 + \alpha
P + \beta P^2 $ with the coefficients $\alpha$, $\beta$ as listed below.}
\label{tab:structure}
\smallskip
\tablesmallfont
%\squeezetable
\begin{ruledtabular}
\begin{tabular}{lllllllll}
    & $V_0$     & $a_0$ & $c_0$  & $B_0$  &$B'$ & $c_0/a_0$ & $\alpha$ & $\beta$ \\
    & (\AA$^3$) & (\AA)     & (\AA)    &  (GPa)   &    &     & (GPa$^{-1}$) & (GPa$^{-2}$)\\
   \colrule
   \AB, Exp.\ (300~K)   & 25.473(1) & 3.0062(1) & 3.2548(1) & 170(1) & 4.8(1) & 1.0827(1) & $-8.8(1) \cdot 10^{-4}$ & $0.77(4) \cdot 10^{-5}$ \\
   \AB, Calc.\ (DFT/GGA) & 25.565    & 2.9977  & 3.2855 & 176.8 & 3.64 & 1.096 & $-12.0 \cdot 10^{-4}$ & $1.09 \cdot 10^{-5}$ \\
   \colrule
   \MB, Exp.\ (300~K)\cite{JM54} & 28.99(1) & 3.0834(3) & 3.5213(6) & 147--155$^\mathrm{a}$ & (4.0)$^\mathrm{a}$\\
\end{tabular}
\end{ruledtabular}
\raggedright $^\mathrm{a}$ References{ } \onlinecite{VSHY01,JHS01,GSGH01}
with the assumption that $B' = 4$.
\end{table*}

\begin{figure}
     \centering
     \includegraphics[width=0.7\hsize,clip]{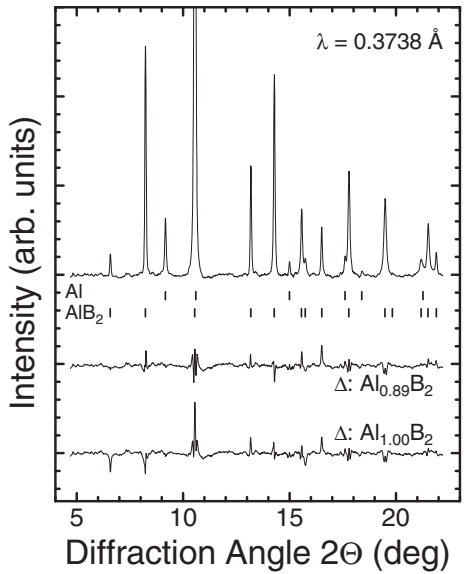}
     \caption{X-ray powder diffraction diagram of \AB\ at 0~GPa
     and 293~K and difference curves
     ($\Delta=I_{\text{exp}}-I_{\text{calc}}$) for
     refinements with and without Al-deficiency. Markers show peak
     positions due to \AB\ and fcc-Al.}
     \label{fig:Al-def}
\end{figure}

Rietveld refinements of the crystal structure with the Al site occupation
as a free parameter indicate that the sample studied here has an Al
deficiency of $\sim$11\%. This is illustrated in Fig.~\ref{fig:Al-def}
where difference curves for refinements of Al$_{1.00}$B$_{2}$ and
Al$_{0.89}$B$_{2}$ are shown together with the experimental diffraction
pattern. A Stephens peak profile \cite{Ste99} was used and a common
isotropic thermal parameter for Al and B was optimized. The weighted
profile $R$ value (without background) reduces from $R_{wp} = 8.5\%$ for
Al$_{1.00}$B$_{2}$ to $R_{wp} = 6.9\%$ for Al$_{0.89}$B$_{2}$. This
indication of substantial Al deficiency in \AB\ is in agreement with
density measurements \cite{MEG64,BGSB00} and recent single-crystal x-ray
diffraction results \cite{BGSB00}.

\subsection{Raman spectra at ambient conditions}
\begin{figure}
     \centering
     \includegraphics[width=0.7\hsize,clip]{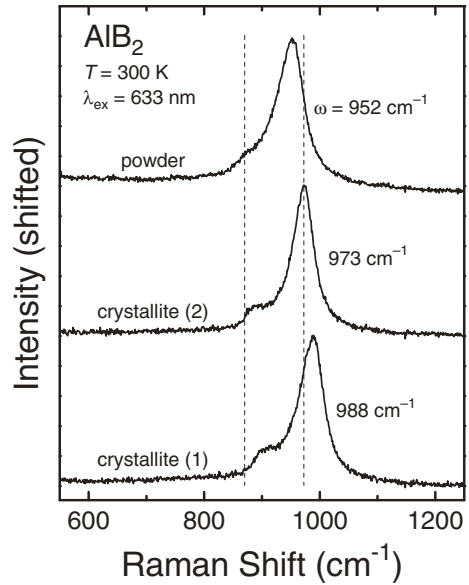}
     \caption{Raman spectra of \AB\ at ambient conditions recorded
     on two crystallites and at a sample spot where no
     crystallites were discernible (`powder').}
     \label{fig:Raman0GPa}
\end{figure}

\AB\ has one Raman-active zone-center phonon mode \cite{BHR01}. It is an
in-plane vibration of the B atoms with $E_{2g}$ symmetry, where
neighboring B atoms move out of phase\cite{KLSK01}. The powder sample we
investigated contained some small shiny crystallites up to $\sim$10~$\mu$m
in size. Figure~\ref{fig:Raman0GPa} shows Raman spectra recorded on two
different crystallites and at a sample spot where no crystallites were
discernible with an optical microscope. Besides a Lorentzian-shaped peak
(FWHM of 40--50~\WN) which was attributed to the $E_{2g}$ mode previously
\cite{BHR01} we observe an additional step-like feature at the
lower-energy side of the main peak. It is clearly visible in the
single-crystal spectra and reduces to a weak shoulder in the powder
spectrum. The powder spectrum resembles that reported by Bohnen \etal\
\cite{BHR01}. The peak position of the $E_{2g}$ mode of the two
crystallites differs by 15~\WN\ ($\omega = 973$ and 988~\WN). In the
powder spectrum the main peak occurs at an even lower energy of 952~\WN.
The step-like feature in the spectra of the crystallites shifts by the
same amount as the main peak indicating that it is intrinsic to \AB. It
appears likely that it is related to a peak in the calculated phonon
density of states\cite{BHR01} which exists slightly below the energy of
the $E_{2g}$ mode.

\subsection{Raman spectra under pressure}
\begin{figure}[bt]
     \centering
     \includegraphics[width=\hsize,clip]{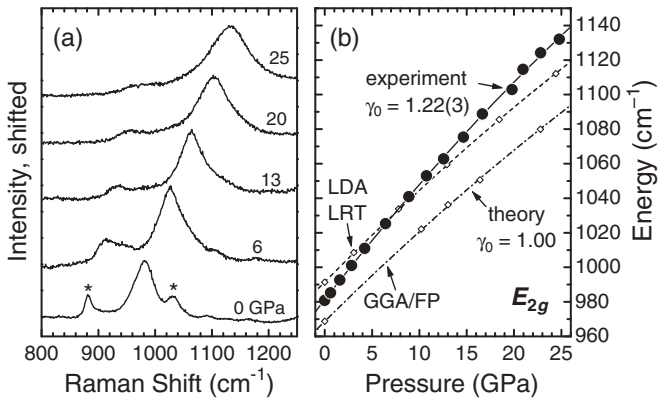}
     \caption{(a) Raman spectra of \AB\ recorded for increasing pressures
     of 0--25~GPa ($T=300$~K). Peaks marked by asterisks are due to the
     methanol/ethanol pressure medium. (b) Experimental and calculated
     energies of the $E_{2g}$ phonon mode. Experimental data are
     represented by circles and a solid line, results of the
     GGA/frozen-phonon and LDA/linear-response-theory calculations by a
     dash-dotted and dashed line, respectively. The lines represent
     quadratic functions fitted to the data.}
     \label{fig:Raman}
\end{figure}

Raman spectra of \AB\ were recorded for increasing pressures up to 25~GPa.
Since the sample is somewhat deformed when pressure is applied and there
seems to be some inhomogeneity of the powder, several spectra were
collected at different locations of the sample. The spectra shown in
Fig.~\ref{fig:Raman}(a) result from averaging about five spectra recorded
at different spots that were selected for a narrow $E_{2g}$ peak and low
background. The zero-pressure frequency of the $E_{2g}$ mode in these
averaged diagrams amounts to 981(1)~\WN. Two additional peaks near 875 and
1025~\WN\ (at 0~GPa) are due to the methanol/ethanol pressure medium.

The $E_{2g}$ phonon frequency increases continuously with increasing
pressure, but with slightly decreasing slope $\d\omega/\d P$. The
zero-pressure mode-Gr{\"u}neisen parameter amounts to $\gamma_0 = 1.22(3)$
(based on $B_0 = 170(1)$~GPa). Up to 10~GPa the  peak width (FWHM) is
essentially constant at $44\pm2$~\WN\ while at higher pressures it
increases to $\sim$90~\WN\ at 25~GPa. This is most likely related to the
solidification of the pressure medium near 10~GPa. From the Raman data
there are no indications for a structural phase transition.

\section{Calculations}
\label{sec:calc}

\subsection{Theoretical Method}

The theoretical methods employed here were described in detail in the
context of our recent work on \MB\ \cite{KLSK01}. In summary,
first-principles calculations of the total energy $E_{tot}$ of the solid are
the basis for the determination of the equation of state (EOS) and of the
phonon frequencies. The energy is computed within the density functional
theory (DFT) using a plane-wave basis and pseudopotentials.

For part of the actual calculations (equation of state and phonon
frequencies in the \emph{frozen-phonon approach}) we employed the VASP
codes \cite{KH93,Kre93,KF96a,KF96b} within the generalized gradient
approximation (GGA) \cite{PW92}. The ultra-soft Vanderbilt-type
pseudopotentials \cite{Van90} were supplied by Kresse and Hafner
\cite{KH94}. The pseudopotential for Al treats explicitly three valence
electrons ($3s^2\, 3p$); no semi-core states are included. A `harder'
variant of the potential was chosen and the nonlinear core correction
\cite{LFC82} is applied in order to improve the transferability. The
calculations are carried out with a plane-wave cutoff energy of 23.6 Ry,
and the Brillouin zone sampling is based on a $\Gamma$-centered $16 \times
16 \times 16$ uniform mesh, yielding 270 to 1170 $k$-points in the
irreducible wedge of the Brillouin zone, depending on the symmetry of the
lattice (equilibrium structure or crystal with the displacements of the
E$_{2g}(\Gamma)$ phonon). As the system is metallic, the $k$-space
integration with the incompletely filled orbitals uses the tetrahedron
method \cite{JA71} with Bl{\"o}chl's corrections \cite{BJA94}.

Phonon frequencies were independently verified using the linear response
theory \cite{Gon97,GL97} as implemented in the ABINIT package
\cite{soft:abinit} within the local density approximation (LDA)
\cite{Sla51,KS65} to the DFT.  We used Hartwigsen-Goedecker-Hutter
pseudopotentials \cite{HGH98} treating Al($3s^2\, 3p$) and B($2s^2\, 2p$)
levels as valence states, a plane-wave cut-off of 60~Ry, and a $12 \times
12 \times 12$ $k$-point mesh. A Gaussian smearing with a broadening
parameter of 0.04~Ry was applied to improve $k$-point sampling in the
special points method. The structural parameters were optimized for each
volume/pressure such that the stress tensor components $\sigma_{xx}$ and
$\sigma_{zz}$ agreed within $10^{-3}$~GPa.

\subsection{Structural Properties}

For fourteen unit cell volumes in the range 21.8--26.2 \AA$^3$ we
calculated (with the VASP codes) the total energy $E(V)$ for different
$c/a$ values, thus determining the optimized structures [$a(V)$, $c(V)$]
and minimized total energies. The latter data were fitted by the Murnaghan
relation for $E(V)$ \cite{Mur44}
\begin{equation}\label{murnE}
E(V) = E_0 + \frac {B_0 V_0}{B'} \left [\frac{V}{V_0} + \frac {
\left ( {V/V_{0}} \right)^{1-B'} - B'}{B'- 1} \right ]
\end{equation}
which provided the static equilibrium volume $V_0$ as well as the bulk
modulus $B_0$ and its pressure derivative $B'$ at zero pressure. The
structural parameters are summarized in Table~\ref{tab:structure}. The
pressure--volume relation, shown in Fig.~\ref{fig:LP-Vol}(b), compares
well with the experimental data. The calculated equilibrium volume is
0.4\% larger than the volume measured at 300~K. The individual lattice
parameters are $-0.3$\% off the experiment for $a_0$, $+0.9$\% for $c_0$,
and consequently +1.2\% for $c_0/a_0$. The variation of $c/a$ under
pressure can be represented by a quadratic function and its coefficients
are given in Table~\ref{tab:structure}.

\subsection{Phonon frequencies}

Energies of the $E_{2g}$ phonon mode as a function of pressure were
initially calculated in the frozen-phonon (FP) approach using the VASP
codes and the GGA. For each phonon and pressure, the atoms are given six
different displacements ranging from $u/a = 0.005$ to 0.035 and the
calculated energy values $E(u)$ are fitted with a quartic polynomial. The
resulting harmonic phonon frequencies and their respective pressure
dependencies are listed in Table~\ref{tab:phonons} together with the
experimental data for the $E_{2g}$ mode and our previous FP-GGA results
for \MB\cite{KLSK01}. As illustrated in Fig.~\ref{fig:Raman}(b) the
calculated zero-pressure frequency is only $\sim$1\% lower than the
experimental value, which is in the typical range for the difference
between experiment and calculations of this type. Regarding the pressure
dependence of the phonon frequency, however, we find an unusually large
deviation of the theoretical results from the experiment. From the
variation of the calculated phonon frequency with volume we obtain a
zero-pressure mode-Gr{\"u}neisen parameter of $\gamma_0 = 1.00$ whereas the
experimental value amounts to $\gamma_0 = 1.22(3)$.

\begin{table*}[tb]
\caption{Pressure and volume dependences of selected phonon frequencies of
\AB\ and \MB. The zero pressure frequency $\omega_0$ and the linear and
quadratic pressure coefficients were obtained by least square fits of
$\omega(P) = \omega_0 + \alpha \cdot P + \beta \cdot P^2$ to the data. The
theoretical mode Gr\"{u}neisen parameters $\gamma_0$ (at the theoretical
equilibrium volume) are derived from a similar quadratic expression for
$\omega(V)$. $P$ has been obtained from $V$ through the calculated
$P$--$V$ relation (see text). The experimental mode Gr\"{u}neisen
parameter is determined from $\omega(P)$ and the experimental $P$--$V$
relation.} \label{tab:phonons}
\smallskip
\begin{ruledtabular}
\begin{tabular}{llllll}
Compound & Mode & $\omega_0$   & $\alpha$  & $\beta$         & $\gamma_0$ \\
         & & (\WN)        & (\WN/GPa)   & (\WN/GPa$^{2}$)   &      \\
\colrule
\AB\  & $E_{2g}$        & 969    & 5.425 & $-0.0235$   & 1.00 \\
      & $E_{2g}$ (exp.) & 981(1) & 7.027 & $-0.0365$   & 1.22(3) \\
      & $B_{1g}$        & 490    & 2.968 & $-0.0087$   & 1.06 \\
\MB\  & $E_{2g}$        & 535    & 8.974 & $-0.0780$   & 2.5 \\
      & $B_{1g}$        & 695    & 3.065 & $-0.0190$   & 0.6 \\
\end{tabular}
\end{ruledtabular}
\end{table*}

Since the deviation of the theoretical results from the experimental data
is larger than usual we performed a second calculation of the $E_{2g}$
frequency under pressure using a rather different approach, namely
linear-response-theory (LRT) in the LDA with the ABINIT package.
Consistent with the common overestimation of bond strengths in the LDA we
obtain here a somewhat larger zero-pressure phonon frequency, such that
the two theoretical values bracket the experimental data at zero pressure.
Over the whole pressure range the LDA/LRT calculation gives phonon
frequencies which are consistently $\sim$2.4\% larger than the
corresponding GGA/FP results. Consequently, we obtain essentially the same
pressure dependence $\d \ln \omega / \d P$ for the two calculations. The
deviation of the calculations from the experimental data is far beyond the
typical uncertainty of such computations. The important difference between
theory and experiment could be that the former is based on the ideal
stoichiometry Al$_{1}$B$_{2}$ whereas the real sample is Al deficient.

For comparison with previous calculations for \MB\ we have also calculated
the frequency $\omega$ of the $B_{1g}$ phonon in \AB\ (out-plane motion of
the boron atoms \cite{KLSK01}), see Table~\ref{tab:phonons}. The pressure
dependence is characterized by a mode-Gr{\"u}neisen parameter $\gamma_0 =
1.06$. In case of a constant, i.e., pressure-independent $\gamma(V) \equiv
\gamma_0$ the relation $\omega(V) = \omega_0 (V/V_0)^{-\gamma_0}$ holds.
Thus, for both the $E_{2g}$ and the $B_{1g}$ mode in \AB\ with $\gamma
\approx 1$ there is a nearly inverse-proportional relation between the
phonon frequency and volume. This is quite different from the situation in
\MB\ where we do not only have the very large $\gamma_0 = 2.5$ for the
$E_{2g}$ mode as noted before but also a rather small $\gamma_0 = 0.6$ for
the $B_{1g}$ phonon (Table~\ref{tab:phonons}).

The frozen phonon calculations also yield information on the anharmonicity
of the phonon modes as described in the context of our \MB\ calculations
\cite{KLSK01}. In essence, the variation of the total energy with atomic
displacement, $E_{\mathrm{tot}}(u)$, can be represented by a polynomial
where the ratio of the quartic to squared quadratic coefficients
$a_4/a_2^2$ is a measure of anharmonicity.  In the harmonic limit $a_4 =
0$. For the $E_{2g}$ mode in \AB\ we obtain $|a_4/a_2^2| < 0.01$~eV$^{-1}$
which is about three orders of magnitude lower than the corresponding
values for \MB\ of $a_4/a_2^2 = 4$--8~eV$^{-1}$ (see
Refs.~\onlinecite{YGLB01,KLSK01}). Small anharmonicities are calculated
for the $B_{1g}$ modes of both compounds: $a_4/a_2^2 = 0.27$ for \AB\ and
$a_4/a_2^2 = -0.05$ for \MB.

\section{Discussion}

\subsection{Structural stability}

In our x-ray diffraction and Raman experiments we do not find any
indication for a structural phase transition or modulation of the
structure. Group-subgroup symmetry considerations \cite{HP01} indicate
numerous possible distortions of the aristotype \AB\ most of which are
realized in intermetallic compounds at ambient pressure. Pressure-induced
structural phase transitions of XY$_2$ intermetallic compounds have not
been studied systematically. In the context of the \AB\ structure a number
of rare-earth metal digallides \cite{SBBS01,SBGS98,BSS02p,SBGS96},
KHg$_{2}$ \cite{BSDT93}, and LaCu$_{2}$ \cite{LHKR00} were investigated at
high pressures. From the available data a transition between the structure
types \AB\ and UHg$_{2}$ appears as a typical route. The KHg$_{2}$
(CeCu$_{2}$) structure type may occur as an intermediate phase. \AB\ and
UHg$_{2}$ are isopointal structures, distinguished only by their $c/a$
ratios. Two clearly separated groups of compounds of \AB\ and UHg$_{2}$
type are observed when plotting the $c/a$ ratio versus ratio of the
metallic radii of XY$_{2}$ intermetallic compounds
\cite{Pea72,Pea79,BSS02p}. The \AB\ and UHg$_{2}$ type branches are
characterized by $c/a$ ratios of 0.95--1.20 and 0.60--0.85, respectively.
The compound \AB\ with $c/a = 1.083$ (at 0~GPa) is located near the center
of the former branch. In the pressure range to 40~GPa explored here it
decreases only to 1.060. Pressures well above 1~Mbar may therefore be
needed for a possible transition towards the UHg$_{2}$ structure. At lower
pressures a transition involving a buckling of the boron honeycomb layers
may occur which could lead to phases of the CeCu$_{2}$, CeCd$_{2}$, or
CaIn$_{2}$ type \cite{HP01,BSS02p}.

\subsection{Raman spectra of \AB\ vs.\ \MB}
Raman spectroscopy is commonly applied to semiconductors and insulators
but only to a much smaller extend to metals. It is essentially the group
of elemental hcp metals that has been studied systematically, already in
the late 1960s at ambient pressure \cite{PFA69} and more recently  at high
pressures (see e.g.\ Refs.~\onlinecite{OJR01,OJ00b} and references
therein). It may therefore be attributed to a lack of reference data that
in case of \MB\ the observation of a very broad Raman feature (FWHM of
$\sim$300~\WN) near 600~\WN\ lead to a still unresolved controversy over
the origin of this peak. It has initially been attributed \cite{BHR01} to
the Raman-active $E_{2g}$ mode which immediately raises the question of
the large linewidth. The large peak width has been related to both strong
electron phonon coupling and to structural disorder. The latter now
appears less likely because Raman spectra of \MB\ powders  are quite
similar to those of recently available small single crystals
\cite{HGPP01,MMRL01pv4} which are presumably less disordered.
High-pressure Raman experiments \cite{KLSK01} have cast doubt on the
assignment to the $E_{2g}$ phonon. They revealed a double-peak structure
with peaks at 603(6)~\WN\ and 750(20)~\WN. Neither of the two peaks could
be attributed to the $E_{2g}$ mode because of severe deviations from
calculated phonon frequencies in terms of zero-pressure frequencies and/or
the pressure dependences.

The present Raman data of \AB\ show that it is possible to obtain Raman
spectra with a well-defined $E_{2g}$ peak from metallic samples of the
\AB\ structure. The difficulties encountered in case of \MB\ are therefore
not likely related to the metallicity of the sample nor intrinsic to the
structure type. Crystallinity also appears to have only a small effect on
the Raman spectrum as the Raman linewidth of the \AB\ powder sample is
comparable to that of the \AB\ crystallites.

There are two properties of \MB\ with regard to phonons which make it
distinct from \AB. First, the whole $E_{2g}$ phonon branch along the
$\Gamma$--$A$ direction in the Brillouin zone exhibits very strong
electron-phonon coupling in \MB\ \cite{KMBA01,AP01,KDJA01,YGLB01}. Second,
the $E_{2g}(\Gamma)$ mode shows pronounced anharmonicity
\cite{YGLB01,KLSK01}. Both electron-phonon and phonon-phonon interaction
 decrease the phonon lifetime and hence increase the phonon
linewidth \cite{RDT02}. They are therefore the most likely causes for the
absence of a well-defined $E_{2g}$ Raman peak in \MB.

\subsection{Metal deficiency in \AB}

Our x-ray diffraction data indicate an Al deficiency of 11\% in \AB\ in
accord with previous density measurements and chemical analysis
\cite{MEG64,BGSB00} as well as recent single-crystal x-ray diffraction
results \cite{BGSB00}. Although the change of the $E_{2g}$ phonon
frequency at different sample spots suggests that there is some variation
of the Al content, there is no indication that growth of aluminum diboride
in the composition Al$_{1.0}$B$_{2}$ is possible. The occurrence of
substantial metal deficiency appears to be common to many (transition)
metal diborides \cite{PLBD96}.

\begin{figure}
     \centering
     \includegraphics[width=\hsize,clip]{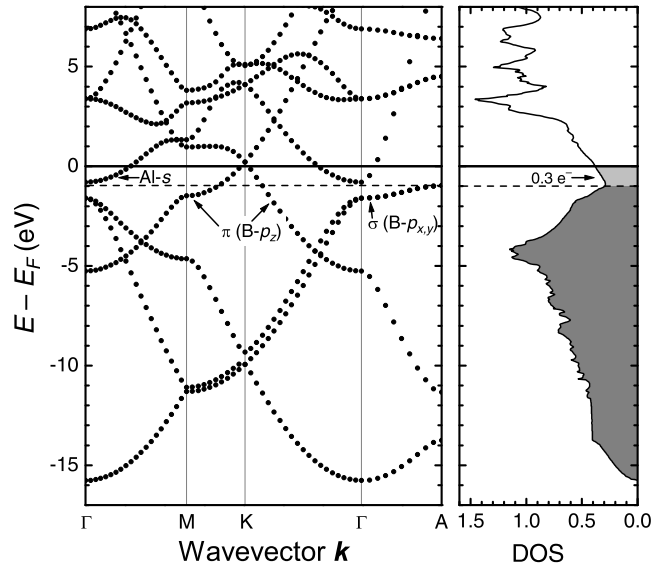}
     \caption{Calculated electronic bandstructure and density of
     states (DOS) of \AB\ at ambient pressure. Energies are given
     with respect to the Fermi energy $E_F$. The DOS is given in
     units of (states/eV/formula unit).}
     \label{fig:BS-DOS}
\end{figure}

In case of \AB\ this metal deficiency has important influence on the
electronic structure. A comparison of the calculated bandstructures of
\AB\ and \MB\ shows that the relative ordering and dispersions of the
bands near the Fermi level are very similar. The difference between \AB\
and \MB\ can largely be treated in a rigid-band picture with a higher band
filling for \AB. It is therefore justified to discuss the observed Al
deficiency of $\sim$10\% in a rigid-band picture, too.
Figure~\ref{fig:BS-DOS} shows the calculated electronic bandstructure
\cite{Wien97:AlB2} and density of states of \AB. In case of the
stoichiometric compound the $\pi$-type bands derived from boron $p_z$
states and the bands with Al-$s$ character (near $\Gamma$) are partially
filled and give rise to the metallic state. 10\% deficiency of Al removes
0.3 valence electrons and consequently lowers $E_F$ by 0.9~eV which leads
to a \emph{complete depletion of the band with \mbox{Al-$s$} character}.
In other words, the experimentally observed Al deficiency is expected to
lead to a qualitative change in the electronic structure compared to the
ideal case of Al$_{1}$B$_{2}$. In terms of the Fermi surface this change
means a removal of the electron pocket around $\Gamma$.

The measured mode-Gr{\"u}neisen parameter of the Al$_{0.9}$B$_{2}$ sample is
$\sim$20\% larger than calculated for Al$_{1.0}$B$_{2}$, a deviation which
is far beyond the typical uncertainty of such computations. The
zero-pressure phonon frequency, on the other hand, seems to be hardly
affected. The apparent difference between experiment and calculation is
the Al deficiency of the sample which was not taken into account in the
theory. We tentatively attribute the discrepancy between the experimental
and calculated mode-Gr{\"u}neisen parameters to the Al-deficiency-induced
electronic changes discussed above. However, it cannot be excluded at this
point that other effects are also at work. A more detailed analysis of the
effect of Al deficiency on the electronic and lattice dynamical properties
of \AB\ is, however, beyond the scope of this work.

\subsection{Metal deficiency in \MB}

Mg deficiency in \MB\ was often indicated by the occurrence of MgO as a
secondary phase in \MB\ samples that were grown from a molar 1:2 mixture
of Mg and B. Variations of \Tc\ and even more of the pressure dependence
of \Tc\ for different samples appeared to be related to non-stoichiometry
of the material. The correlation between composition, structural
parameters and \Tc\ has been established in an experiment by Indenbom
\etal\ \cite{IUKB01}. By diffusion of Mg into a boron cylinder they
produced a sample with a composition changing gradually between
Mg$_{1.0}$B$_{2}$ and Mg$_{0.8}$B$_{2}$. With decreasing Mg content the
lattice parameter $c$ increased by 0.003~{\AA} (0.1\%) and \Tc\ increased from
37.2 to 39.0~K.

Tissen \etal\ \cite{TNKK01} furthermore pointed out a correlation between
the zero-pressure critical temperature $T_{c,0}$ and the pressure
derivative $\d\Tc/\d P$: As $T_{c,0}$ increases from 37.3 to 39.2~K for
various samples, $\d\Tc/\d P$ changes from $-2.0$ to $-1.1$~K/GPa.
Monteverde \etal\ \cite{MNRR01} discussed a similar observation on a
smaller number of samples in terms of the electronic band structure and
band-filling effects related to the Mg non-stoichiometry. On the other
hand, the pressure-induced changes of the electronic density of states
calculated for \MB\ are too small to account for the observed decrease of
\Tc\ under compression \cite{LS01}. It is rather the increase in the
relevant phonon frequencies which provides the main contribution to the
pressure dependence of \Tc\ \cite{LS01,VSHY01,CZH02}. It would therefore
be rather surprising if electronic density effects -- i.e., the electronic
density $N(E_F)$ at the Fermi level -- were responsible for the large
sensitivity of $\d\Tc/\d P$ on Mg non-stoichiometry.

The indications that Al deficiency in \AB\ may affect the pressure
dependence of the $E_{2g}$ phonon frequency hints at an alternative
possible explanation for the large sensitivity of $\d\Tc/\d P$ on Mg
deficiency in \MB. It was first pointed out by Yilderim \etal\
\cite{YGLB01} that the $E_{2g}$ phonon mode in \MB\ exhibits a very large
anharmonicity. Boeri \etal\ \cite{BBCP01pv1} showed theoretically that
this effect arises in \MB\ because here the Fermi level $E_F$ is located
only $\sim$0.5~eV below the top of $\sigma$ bands of the equilibrium
structure. The lattice distortion of the $E_{2g}$ mode induces a splitting
of these $\sigma$ bands large enough that the lower split-off band sinks
completely below $E_F$ \cite{AP01,BBCP01pv1}. This does not happen in \AB\
and graphite, and anharmonicity is indeed negligible.

It is also noteworthy that the $E_{2g}$ mode in \AB\ is much higher in
energy than the $B_{1g}$ phonon whereas the reversed order is calculated
for \MB\ (see Table~\ref{tab:phonons}) although both compounds are
structurally quite similar. This effect was pointed out before and studied
in Mg$_{1-x}$Al$_{x}$B$_{2}$ mixed crystals by Renker \etal\ \cite{RBH02}.
The interchange, which occurs only in undoped or moderately substituted
material ($0<x<0.2$), was also attributed to the electronic changes,
especially the disappearance of the hole pockets from the Fermi surface
for $x > 0.2$.

The metal content in \MB\ affects the band filling, a larger Mg deficiency
moving the Fermi level further below the top of the $\sigma$~bands. It is
therefore to be expected that the anharmonicity of the $E_{2g}$ mode
should decrease with decreasing Mg content. Lattice dynamical calculations
showed that the $E_{2g}$ anharmonicity decreases with increasing pressure
\cite{KLSK01} and the initially very large mode-Gr{\"u}neisen parameter
decreases too \cite{KLSK01}. If the mode-Gr{\"u}neisen parameters decreases as
function of band filling at ambient pressure, i.e., due to
non-stoichiometry of \MB, it would, qualitatively, lead to the observed
relation between $\d\Tc/\d P$ and  Mg deficiency. This effect would be a
manifestation of the changes of the lattice dynamics rather than changes
of the electronic density of states. A more detailed and quantitative
analysis is certainly needed, but the present results are indication of
the importance of stoichiometry with regard to the superconducting
properties of \MB, specifically the pressure dependence of \Tc.

\section{Conclusions}

We have studied the crystal structure of \AB\ by x-ray powder diffraction
to 40~GPa. The compressibility is moderately anisotropic consistent with
the anisotropic bonding properties. In the pressure range studied here we
did not observe a structural phase transition. Our x-ray diffraction data
indicate an Al deficiency of $\sim$11\% in agreement with previous
reports. Despite the neglect of this non-stoichiometry in our
first-principles calculations, the calculated structural properties are in
good agreement with the experiment.

The $E_{2g}$ zone-center phonon in metallic \AB\ can be observed as a
well-defined Raman peak. We conclude that the lack of such a Raman feature
in \MB\ is neither related to the metallicity or disorder of the sample
nor is it a generic property of \AB-type compounds. Our observations
rather support the view that it is due to the strong electron-phonon
coupling and/or anharmonicity which are distinct properties of \MB. We
found some deviation of the calculated pressure dependence of the $E_{2g}$
phonon frequency of \AB\ from the experimental data and tentatively
attributed this to the Al deficiency of the \AB\ sample which was not
taken into account in the theory.

Correlations between non-stoichiometry of \MB\ and its superconducting
properties have been pointed out previously. Here we considered possible
effects of Mg deficiency in \MB\ on its electronic structure and lattice
dynamics. The anticipated changes are consistent with the available
experimental data on the correlation between Mg content and the pressure
dependence of \Tc. This leads us to propose that the large variation of
the pressure dependence of \Tc\ ($-0.7$ to $-2.0$ K/GPa) in \MB\ in
various experiments may be caused by the effect of non-stoichiometry on
the lattice dynamics, mediated via changes in the electronic structure of
\MB.

\begin{acknowledgments}
We thank Yu.~Grin for calling our attention to the issue of the
non-stoichiometry of \AB. The computer resources used in this work were in
part provided by the Scientific Committee of IDRIS, Orsay (France).
\end{acknowledgments}

%\clearpage

% ----------------------------------------------------------------

%\bibliographystyle{prsty}
%\bibliography{papers,misc,AlB2}

\begin{thebibliography}{10}
\bibitem{Wien97:AlB2}
The electronic bandstructure and density of states were calculated using the
  full-potential linearized augmented-plane-wave method method as implemented
  in the WIEN97 code \cite{soft:WIEN97}. For the exchange-correlation potential
  we employed the generalized gradient approximation of
  Ref.~\onlinecite{PBE96}. Scalar-relativistic corrections were included. The
  Al $2p$ states were treated as band states using the local orbital extension
  of the LAPW method \cite{soft:WIEN97,Sin91}. For $k$-point sampling employing
  the tetrahedron method, 296 points were used in the irreducible wedge of the
  Brillouin zone (4536 in total); other parameters were $R_{\rm MT} \cdot
  K_{\rm max}=8.0$; $R_{\rm MT}=1.8$ and 1.4 $a_B$ for Al and B,
  respectively; $l_{\rm max}=10$; $G_{\rm max}=15$.
\end{thebibliography}

% ----------------------------------------------------------------

\end{document}